# MoS$_2$ Nanoribbon Transistors: Transition from Depletion-mode to Enhancement-mode by Channel Width Trimming

Han Liu, Jiangjiang Gu, *Student Member, IEEE,* and Peide D. Ye, *Senior Member, IEEE*

*Abstract* - We study the channel width scaling of back-gated MoS$_2$ metal-oxide-semiconductor field-effect transistors (MOSFETs) from 2 μm down to 60 nm. We reveal that the channel conductance scales linearly with channel width, indicating no evident edge damage for MoS$_2$ nanoribbons with widths down to 60 nm as defined by plasma dry etching. However, these transistors show a strong positive threshold voltage (V$_T$) shift with narrow channel widths of less than 200 nm. Our results also show that transistors with thinner channel thicknesses have larger V$_T$ shifts associated with width scaling. Devices fabricated on a 6 nm thick MoS$_2$ crystal underwent the transition from depletion-mode to enhancement-mode.

## I. INTRODUCTION

THE triumph of aggressive scaling of silicon based integrated circuits has dramatically changed our lifestyle in the past couples of decades. However, as the scaling of silicon approaches its physical limit, efforts in finding alternative channel materials have been made for the extension of the Moore's Law. Of these materials, Ge and III-V materials are among the most promising candidates because of their high carrier mobilities. They have been widely studied for logic applications in the past years [1-4]. Although graphene, a single layer of carbon atoms having superior carrier mobilities of up to 200,000 cm$^2$/V·s, has been recognized as another material candidate, its gapless nature limits its further application in logic devices [5]. Nevertheless, the discovery of graphene has spurred research of other two-dimensional (2D) layered structures, including boron nitride, topological insulators (Bi$_2$Te$_3$, Bi$_2$Se$_3$, etc.), and transitional metal dichalcoginides (TMDs) [6-10]. TMDs, e.g. MoS$_2$, have enjoyed several advantages in device applications because they have large bandgaps (usually >1 eV), satisfactory electron mobilities of up to several hundred, good thermal stability, and can be used to form ultrathin body transistors with atomic layers, which make them a desirable channel material with superior immunity to short channel effects [11-13]. The MoS$_2$ devices are mostly n-type transistors in experimental observations, which might be attributed to the stoichiometric composition [10,15,16]. Also, its charge neutrality level is in the vicinity of the conduction band (E$_C$), thus n-type contacts can be more easily made for MoS$_2$ transistors and the transistors behave as depletion-mode n-channel MOSFETs with large negative threshold voltages [14]. The 2D nature of MoS$_2$ (as well as other layered materials) makes it difficult to realize channel doping as a way to achieve desirable positive V$_T$ or enhancement-mode operation for various desirable circuitry configurations. Therefore, making an enhancement-mode MoS$_2$ MOSFET is quite challenging.

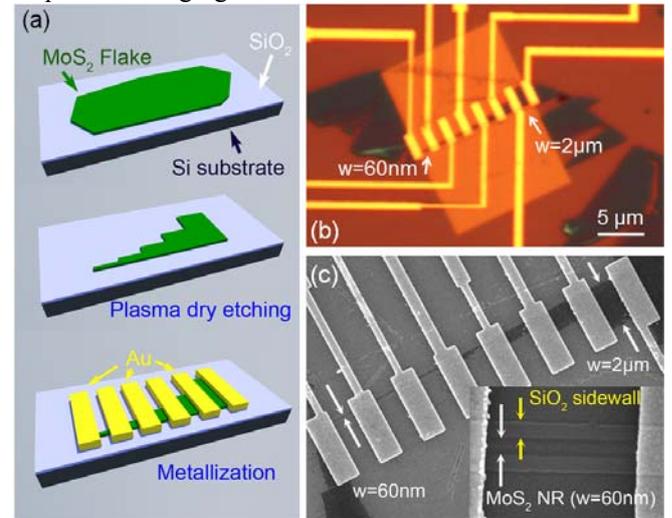

Fig 1. (a) Schematic illustrations of device fabrication. (b) Optical image of one set of transistors (D1) fabricated on a 6 nm thick MoS$_2$ crystal. (Scale bar: 5 μm) (c) SEM image of another set of devices (D2). (Inset: magnified image of the 60 nm wide MoS$_2$ nanoribbon transistor.)

## II. EXPERIMENTS

In this letter, we study the width scaling of MoS$_2$ transistors by forming nanoribbon channels, and show that the V$_T$ of MoS$_2$ transistors can be modulated to be both positive and negative through appropriate width selection. The fabrication process of sets of MoS$_2$ transistors is shown in Figure 1(a), as described below. MoS$_2$ flakes were mechanically exfoliated from bulk ingot (SPI Supplies) and then transferred to a heavily p-doped silicon substrate (0.01-0.02 Ω·cm) with a 300 nm SiO$_2$ capping layer. The silicon substrate served as a global back gate, while the 300 nm SiO$_2$ served as the gate dielectric. After the flake transfer, we used electron beam lithography to pattern the flake, followed by plasma dry etching (BCl$_3$: 15 sccm, Ar: 60 sccm, Pressure: 0.6 Pa, RF source power: 100W, RF Bias Power: 50W, time: 5 min) to remove the excess parts of the flakes, leaving connected rectangles with a fixed length (2 μm) but various widths to be used as device channels. The widths of these rectangles were varied from 2 μm down to 60 nm. Finally, contacts were defined by electron beam lithography, followed by a 50 nm Au metallization by electron beam deposition. The

Manuscript received on April 13, 2012. This work was supported in part by National Science Foundation (Grant No.CMMI-1120577). The authors would like to thank Nathan Conrad for critical reading of the manuscript.

The authors are with the School of Electrical and Computer Engineering and the Birck Nanotechnology Center, Purdue University, West Lafayette, IN 47907 USA (Tel: +1 (765) 494-7611, Fax: +1 (765) 496-7443, E-mail: yep@purdue.edu).

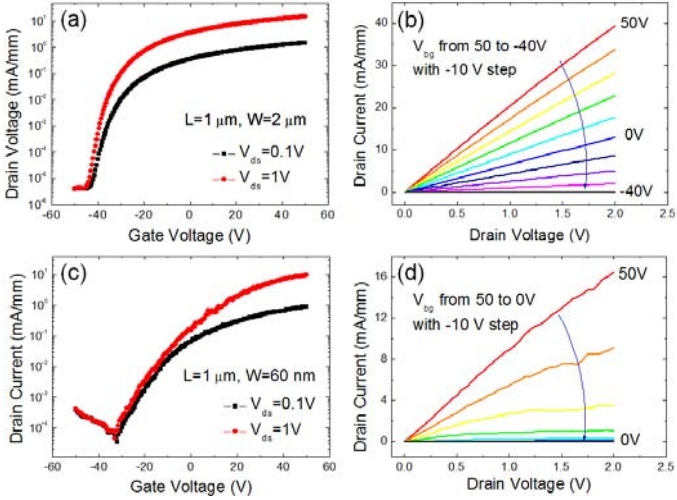

Fig. 2. (a) (b) Transfer and output curves of a transistor with 2 μm channel width on a 6 nm thick $MoS_2$ crystal (D1s). (c) (d) Transfer and output curves of the transistor with 60 nm channel width on the same crystal.

contact bars were 1 μm wide, centered on the edge between two neighboring channel areas. The final set of devices have a fixed channel length of 1 μm and widths of 2 μm, 1 μm, 500 nm, 200 nm, 100 nm, 80 nm, and 60 nm. Three sets of devices were fabricated on three large flakes with thicknesses of 6, 6 and 11 nm, as determined by atomic force microscopy (AFM). The optical image and scanning electron microscopy (SEM) image of these sets of devices are shown in Figures 1(b) and 1(c). Over-etching of the $MoS_2$, in order to guarantee complete removal of excess $MoS_2$ crystals, created a rectangular step in the $SiO_2$ capped substrate surrounding the flake, and also created $SiO_2$ sidewalls at the edges of the $MoS_2$ channels (Figure 1(c) inset). Smooth edges without obvious damage by dry etching was observed at the $MoS_2$ nanoribbons.

Figure 2(a)-(d) show the transfer and output characteristics of the devices with 2 μm and 60 nm widths, selected from one of three sets of devices. These devices were fabricated using one of the 6 nm thick crystals (D1). Figure 2(a) shows well-behaved transfer curves from the 2 μm wide transistor. The current on/off ratio is approximately $10^7$, as the ultrathin $MoS_2$ crystal can be easily depleted at negative gate biases. The extrinsic field-effect mobility of this device is 21.8 $cm^2/V·s$, which can be further improved by high-k dielectric passivation [10,15]. The transfer curve of the 60 nm wide nanoribbon device is shown in Figure 2(c). Compared to Figure 2(a), the transfer curves are noisy, due to fewer conduction modes in the nanoribbons, as well as an increased leakage current at negative bias. This increased gate leakage could be ascribed as more defects are induced in the vicinity of etching windows in the $SiO_2$. We also observe a larger subthreshold swing (SS). The SS for the 2 μm wide device is around 2 V/dec, however for the 60 nm wide device, this value increases to almost 10 V/dec. The SS value of the 2 μm wide device indicates a reasonably good interface (interface trap density $D_{it}$ ~ $2.3×10^{12}/cm^2·eV$) between the $MoS_2$ crystal and $SiO_2$ dielectric. If we replace the 300 nm $SiO_2$ layer with 5 nm of $Al_2O_3$, while assuming that $D_{it}$ remains unchanged, the SS would be significantly reduced to ~75 mV/dec by simply applying $SS=kT/q(1+C_{it}/C_{ox})$, where $k$ is is the Boltzmann constant, $T$ is the temperature and $C_{ox}$ and $C_{it}$ are capacitances modeling the oxide and interface traps, respectively. The differences in SS for the two devices are expected because edge roughness and defects play more important roles in nanoribbon transistors. We also observe a large difference in threshold voltages for these two devices. At a 2 V drain voltage with zero gate bias, the drain current for the 2 μm wide device is 13.0 mA/mm, showing typical depletion-mode operation. However, for the narrower device, the drain current is near zero with zero gate bias, indicating an obvious $V_T$ shift to the positive side, signaling enhancement-mode operation. At the same $V_{ds}$=2V and $V_{gs}$=50V, we achieve the highest drain current density to be 39.4 mA/mm for the 2 μm wide device and 16.4 mA/mm for the 60 nm wide device. This difference in normalized drain current suggests that $V_T$ is different for these two devices, assuming that the current scales linearly with the channel width, which is verified below.

In order to confirm that the current scales linearly with various channel width, we plot the total conductance (1 over on-resistance) versus channel width for all three sets of devices, as shown in Figure 3(a). The on-resistance ($R_{on}$) of the transistor has contributions from the contact resistance and channel resistance. Since the contact area and width for all transistors scales with channel width, the contact resistance should scale linearly with the channel width. The same is true for the channel resistance. After determining $V_T$ via linear extrapolation from transfer characteristics, $R_{on}$ is extracted at the same reference voltage point $V_{gs}$=$V_T$+26V. Our result shows that the $R_{on}$ or total conductance scales linearly with channel width. As expected, $MoS_2$ acts as a conventional semiconductor, in great contrast to 3D topological insulators such as $Bi_2Te_3$ or $Bi_2Se_3$, where edge or surface conductance could be enhanced or dominate [17]. The width dependent conductance is about 6-7 μS/μm for the two 6 nm thick devices (D1 and D2) and 5 μS/μm for the 11 nm device (D3). We are not clear about why the thicker device has a lower conductance and this result needs further investigation. As shown in the inset of Figure 3(a), with much reduced channel widths, the extracted conductance becomes noisier as fewer conduction modes are available in the nanoribbons. But, the data points still fall along the scaling trend.

Lastly, we studied the $V_T$ shift of all devices associated with the channel width. The $V_T$ is extracted from the linear extrapolation method at a low drain voltage. $V_T$ is calculated by $V_T = V_{GSi} – V_{ds}/2$, where $V_{GSi}$ is the intercept gate voltage, and $V_{ds}$ is the drain voltage [18]. Threshold voltages for all three sets of devices are plotted in Figure 3(b). Similar trends can be observed for all sets of devices. The $V_T$ remains constant for transistors with wider channel width (W>500 nm). As the width of the channel is narrowed down to 200 nm, we start to observe the $V_T$ shift to positive values. Apparently, transistors with thinner bodies (D1 and D2) are more likely to be influence by this effect. For one of the 6 nm thick set of transistors (D2), the threshold voltage ultimately shifts from -20V to 30V, indicating a clear transition from being a depletion-mode transistor to being an enhancement-mode operation just by trimming down the channel width. The geometry of these nanoribbon transistors with channel width less than 100 nm has a similar structure to Si FinFETs, if we ignore that the $MoS_2$ channel is modulated only from the back gate. Similar trends of $V_T$ shift have also been observed in Si FinFETs as well as InGaAs nanowire transistors. This narrow channel effect was ascribed to the lateral expansion of depletion layer due to fringing field effect or quantum confinement in device channels [19-21]. However, the channel widths are strickly defined in our $MoS_2$ transistors thus they cannot have a lateral expansion in depletion layer. Also, they are much wider than those of these Si FinFETs and InGaAs nanowire MOSFETs. We believe that our $V_T$ shift is due to edge depletion, similar to what has been observed in majority carrier GaN nanoribbon devices [22]. The edge depletion could be induced by either electric fields or ambient molecules (e.g. $H_2O$) adsorbed at the $MoS_2$ surface. Given our previous surface study of atomic layer deposition (ALD) growth on 2D crystals, these polarized molecules can be strongly adsorbed at $MoS_2$ surface and even persist at 300-400 ºC [23]. As expected, the $V_T$ of the devices fabricated on the thicker crystal (D3) show relatively minor shifts compared to the devices with thinner flakes, as shown in the same figure. The observation of $V_T$ shifts for $MoS_2$ transistors with width scaling is important. The 2D nature of $MoS_2$ and other TMD based transistors makes them difficult to engineer the channel through doping. This demonstrated approach, using the width to achieve $V_T$ adjustments on the same starting channel material in order to realize both enhancement-mode and depletion-mode operation, is a simple and favorable method for circuit designs such as to realize an enhancement-mode/depletion-mode based inverter.

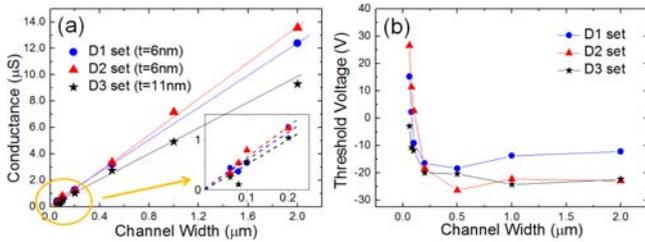

Fig. 3. (a) Extracted channel conductance of all sets of devices versus channel width. The dashed lines in the inset are guided by eye. (b) Extracted threshold voltage of all sets of devices versus channel width.

### III. CONCLUSION

In summary, we have studied effect of the width scaling on $MoS_2$ transistors. We demonstrated that the channel conductance scales linearly for sets of devices with various channel widths down to 60nm. We also revealed that the threshold voltage has a significant shift from negative to the positive for transistors with channel widths of less than 200 nm. As a result, these transistors show a clear transition from depletion-mode to enhancement-mode operation simply by channel width trimming. Our result provides a new approach for threshold voltage engineering and is favorable for circuit applications based on 2D crystal nanomaterials.

REFERENCES

[1] R. Pillarisetty, "Academic and industry research progress in germanium nanodevices", *Nature,* vol. 479, pp. 324-328, 2011; references therein.
[2] Y. Xuan, Y. Q. Wu, and P. D. Ye, "High-Performance Inversion-Type Enhancement-Mode InGaAs MOSFET With Maximum Drain Current Exceeding 1 A/mm", *IEEE Electron Devices Letters,* vol. 29, pp. 294-296, 2008.
[3] H. Ko, K. Takei, R. Kapadia, S. Chuang, H. Fang, P. W. Leu, K. Ganapathi, E. Plis, H. S. Kim, S.-Y. Chen, M. Madsen, A. C. Ford, Y.-L. Chueh, S. Krishna, S. Salahuddin and A. Javey, "Ultrathin compound semiconductor on insulator layers for high performance nanoscale transistors", *Nature,* vol. 468, pp. 286-289, 2010.
[4] M. Xu, R. Wang, and P. D. Ye, "GaSb Inversion-Mode PMOSFETs With Atomic-Layer-Deposited $Al_2O_3$ as Gate Dielectric", *IEEE Electron Devices Letters,* vol. 32, pp.883-885, 2011.
[5] Y. Zhang, J. W. Tan, H. L. Stormer, P. Kim, "Experimental observation of the quantum Hall effect and Berry's phase in graphene", *Nature,* vol. 438, pp. 201-204, 2005.
[6] K. S. Novoselov, D. Jiang, F. Schedin, T. J. Booth, V. V. Khotkevich, S. V. Morozov, and A. K. Geim, "Two-dimensional atomic crystals", *Proc. Natl Acad. Sci. USA,* vol. 102, pp. 10451-12453, 2005.
[7] H. Zhang, C. X. Liu, X. L. Qi, X. Dai, Z. Fang, and S. C. Zhang, "Topological insulators in $Bi_2Se_3$, $Bi_2Te_3$ and $Sb_2Te_3$ with a single Dirac cone on the surface", *Nature Physics,* vol. 5, pp. 438-442, 2009.
[8] H. Liu and P. D. Ye, "Atomic-layer-deposited $Al_2O_3$ on $Bi_2Te_3$ for topological insulator field-effect transistors", *Appl. Phys. Lett.,* vol. 99, 052108, 2011.
[9] V. Podzorov, M. E. Gershenson, Ch. Kloc, R. Zeis, and E. Bucher, "High-mobility field-effect transistors based on transition metal dichalcogenides", *Appl. Phys. Lett.,* vol.84, 3301, 2004.
[10] B. Radisavljevic, A. Radenovic, J. Brivio, V. Giacometti and A. Kis, "Single-layer $MoS_2$ transistors", *Nature Nanotechnology,* vol.6, pp. 147-150, 2011.
[11] K. F. Mak, C. Lee, J. Hone, J. Shan, and T. F. Heinz, "Atomically Thin $MoS_2$: A New Direct-Gap Semiconductor", *Phys. Rev. Lett.,* vol. 105, 136805, 2010.
[12] Y. Yoon, K. Ganapathi, and S. Salahuddin, "How Good Can Monolayer $MoS_2$ Transistors Be?", *Nano Lett.,* vol. 11, pp. 3768-3773, 2011.
[13] H. Liu, A. T. Neal and P. D. Ye, "Channel Length Scaling of $MoS_2$ MOSFETs", submitted to *ACS Nano,* 2012.
[14] B. L. Abrams and J. P. Wilcoxon, "Nanosize semiconductors for photooxidation," *Crit. Rev. Solid State Mater. Sci.*, vol. 30, pp. 153-182, 2005.
[15] H. Liu and P. D. Ye, "$MoS_2$ Dual-Gate MOSFET with Atomic Layer Deposited $Al_2O_3$ as Top-Gate Dielectric", *IEEE Electron Devices Letters,* vol. 33, pp. 546-548, 2012.
[16] Z. Yin, H. Li, H. Li, L. Jiang, Y.Shi, Y. Sun, G. Lu, Q. Zhang, X. Chen, H. Zhang, "Single-Layer $MoS_2$ Phototransistors", *ACS Nano,* vol. 6, pp. 74-80, 2012,
[17] J. G. Analytis, R. D. McDonald, S. C. Riggs, J.-H. Chu, G. S. Boebinger and I. R. Fisher, "Two-dimensional surface state in the quantum limit of a topological insulator", *Nature Physics,* vol. 6, pp. 960-964, 2010.
[18] D. K. Schroder, "Semiconductor Material and Device Characterization", 3rd Edition, Wiley, New York (2006)
[19] K. E. Kroell and G. K. Ackermann, "Threshold Voltage of Narrow Channel Field Effect Transistors", *Solid-State Electrons,* vol. 19, pp. 77-81, 1976.
[20] H. Lee, L.-E. Yu, S.-W. Ryu, J.-W. Han, K. Jeon, D. Y. Jang, K.-H. Kim, J. Lee, J.-H. Kim, S. C. Jeon, G. S. Lee, J. S. Oh, Y. C. Park , W. H. Bae, H. M. Lee, J. M. Yang, J. J. Yoo, S. I. Kim, Y.-K. Choi, "Sub-5 nm all around gate FinFET for ultimate scaling," in *VLSI Symp. Tech. Dig.,* 2006, pp. 58-59.
[21] J.J. Gu, Y.Q. Liu, Y.Q. Wu, R. Colby, R.G. Gordon, and P.D. Ye, "First Experimental Demonstration of Gate-all-around III-V MOSFETs by Top-down Approach", *IEDM Tech. Dig*. pp. 769-772, 2011.
[22] B. Lu, E. Matioli and T. Palacios, "Tri-Gate Normally-off GaN Power MISFET", *IEEE Electron Device Letters,* vol. 33, pp. 360-362, 2012.
[23] H. Liu, K. Xu, X. J. Zhang and P. D. Ye, "The Integration of High-k dielectric on Two-Dimensional Crystals by Atomic Layer Deposition", *Appl. Phys. Lett.,* vol.100, 152115, 2012.